\documentclass[aps,prl,amsmath,amssymb,superscriptaddress,twocolumn,longbibliography]{revtex4-2}
\usepackage{bbm}
\usepackage{graphicx}
\usepackage{dcolumn}
\usepackage{bm}
\usepackage{subfigure}
\usepackage{amsmath}
\usepackage{stmaryrd}
\usepackage{times}
\usepackage{graphicx}
\usepackage{dcolumn}
\usepackage{color}
\usepackage{makecell}
\usepackage{tabularx,multirow}
\usepackage{appendix}
\usepackage{amssymb}
\usepackage{pifont}
\usepackage[colorlinks,linkcolor=blue,citecolor=blue,urlcolor=black]{hyperref}
\newcommand{\cmark}{\ding{51}}%
\newcommand{\xmark}{\ding{55}}%

\begin{document}
\title{Electric-Field Switchable Magnetic Spin Hall Effect}
\author{Mingbo Dou}
\affiliation{School of Physics, Harbin Institute of Technology, Harbin 150001, China}
\author{Xu Chen}
\affiliation{School of Physics, Harbin Institute of Technology, Harbin 150001, China}
\author{Qin Zhang}
\affiliation{School of Physics, Harbin Institute of Technology, Harbin 150001, China}
\author{Xianjie Wang}
\affiliation{School of Physics, Harbin Institute of Technology, Harbin 150001, China}
\affiliation{Heilongjiang Provincial Key Laboratory of Advanced Quantum Functional Materials and Sensor Devices, Harbin 150001, China}
\author{Xue-Zeng Lu}
\affiliation{Key Laboratory of Quantum Materials and Devices of Ministry of Education, School of Physics, Southeast University, Nanjing 211189, China}
\author{Jia Zhang}
\affiliation{School of Physics and Wuhan National High Magnetic Field Center, Huazhong University of Science and Technology, 430074 Wuhan, China}
\author{L. L. Tao}
\affiliation{School of Physics, Harbin Institute of Technology, Harbin 150001, China}
\affiliation{Heilongjiang Provincial Key Laboratory of Advanced Quantum Functional Materials and Sensor Devices, Harbin 150001, China}
\date{\today}
\begin{abstract}
It is established that the polarity of a time-reversal-odd ($\mathcal{T}$-odd) physical quantity can be reversed under the $\mathcal{T}$ operation. Here, we use the spin-group analysis to directly demonstrate that the $\mathcal{T}$-odd magnetic spin Hall effect in ferroelectric altermagnets can be switchable by electric fields beyond the $\mathcal{T}$ operation. This arises from the ferroelectric switching of the nonrelativistic spin splitting, which swaps the roles of spin up and down channels in the reciprocal space. As a result, the $\mathcal{T}$-odd spin conductivity that are proportional to the spin-polarized conductivity difference reverses its polarity upon polarization switching. We identify spin-group operations to switch both the polarization and the magnetic spin Hall effect simultaneously for non-centrosymmetric spin point groups. Then, we exemplify those phenomena in the ferroelectric altermagnet VOI$_2$ monolayer based on density functional theory calculations and an effective Hamiltonian analysis. Our findings not only provide novel strategies to switch the magnetic spin Hall effect using the dissipation-free electric field but also open a promising avenue for electrically programmable spintronic devices.
\end{abstract}
\maketitle
It is known that the polarity of a magnetic quantity (e.g., magnetic field, magnetic moment, and anomalous Hall effect) would be reversed under the time-reversal (abbreviated as $\mathcal{T}$) operation\cite{callaway}. A switchable magnetic quantity is crucial for magnetic information storages, e.g., the switchable magnetization for magnetic tunnel junctions\cite{nm862,nm868} and the switchable Néel vector (antiferromagnetic order) for antiferromagnetic tunnel junctions\cite{nc7061,prx011028,prl197201,na485,na490}. The conventional way to switch a magnetic quantity is using an external magnetic field or the electric current induced spin-transfer torque\cite{nm372} or spin-orbit torque\cite{rmp035004,apl120502,pms100761,apr011305,npj8}. This is legitimate since both the magnetic field and the electric current are $\mathcal{T}$-odd quantities, which are compatible with the $\mathcal{T}$-odd magnetism. Of particular interest for ultra-low power spintronic devices is using the dissipation-free electric field to switch a magnetic quantity\cite{pms100761,npj8,apr011305,nm203,na35,prx021033,prl096803}. However, it is not straightforward to control or switch a magnetic quantity using electric fields due to the fundamental distinct $\mathcal{T}$-odd magnetism and $\mathcal{T}$-even electric field. 

On the other hand, the charge-to-spin conversion is of vital importance for the generation of spin currents\cite{na509,npj27,nrm258}. A charge-to-spin conversion is characterized by a spin conductivity tensor $\sigma_{jk}^{i}$, which describes a charge current along the $k$ direction inducing a spin current along the $j$ direction with the spin polarization along the $i$ direction, where the indices $i, j, k=x, y, z$ denote Cartesian components. In general, $\sigma_{jk}^{i}$ consists of two distinct terms\cite{prb174423,prl187204,na627,prb224401,npj46}, that is, $\sigma_{jk}^{i}=\sigma_{jk}^{i, \text{odd}}+\sigma_{jk}^{i, \text{even}}$,  where $\sigma_{jk}^{i, \text{odd}}$ is odd under $\mathcal{T}$ but $\sigma_{jk}^{i, \text{even}}$ is even under $\mathcal{T}$. The $\mathcal{T}$-odd contribution $\sigma_{jk}^{i, \text{odd}}$ represents an unconventional spin response, which is also called the magnetic spin Hall effect (SHE) absent in nonmagnetic materials. This was firstly explored in the non-collinear antiferromagnets Mn$_3$Sn and Mn$_3$Ir based on density functional theory (DFT) calculations\cite{prl187204} and was later experimentally observed in Mn$_3$Sn\cite{na627}, for which the magnetic SHE is reversed upon the triangularly ordered moments switching ($\mathcal{T}$ operation). Intriguingly, it was demonstrated that the magnetic SHE generated by the Mn$_3$Sn can exert an efficient anti-damping torque, which enables the deterministic switching of the perpendicular magnetization of the adjacent Ni/Co multilayer\cite{nc4447}. In contrast, the $\mathcal{T}$-even contribution $\sigma_{jk}^{i, \text{even}}$ represents the conventional SHE whose sign remains invariant under magnetic moments switching and the spin-orbit coupling (SOC) is essential\cite{rmd1213,prl246602}. Recently, the research of the charge-to-spin conversion contributed by the magnetic SHE has been reinvigorated and blooming in burgeoning altermagnets\cite{prx031042,prx040501,afm2409327,jcas2257,nrm2025,nature837}. As distinct from the conventional ferromagnets and antiferromagnets, altermagnets are characterized by the zero net magnetization and the sizable nonrelativistic spin splitting of the order of $1.0$ eV induced by the exchange coupling\cite{jpsj123702,prb014422,nc2846}. In particular, a promising spin-splitter torque was proposed based on the nonrelativistic charge-to-spin conversion dictated by $\sigma_{jk}^{i, \text{odd}}$\cite{prl127701,prl197202,prl137201,nc5646}. A remarkable field-free switching of the perpendicular magnetization driven by the spin-splitter torque exerted by $\sigma_{jk}^{i, \text{odd}}$ from the altermagnetic RuO$_2$ has been achieved at room temperature\cite{prl137201,nc5646,nc1309}.

It is desirable to achieve the electric-field switchable magnetic SHE towards the ultra-low power and programmable spintronic devices. However, this still remains elusive due to the fundamental inconsistence between the $\mathcal{T}$-odd magnetic SHE and the $\mathcal{T}$-even electric field. In this work, we propose an efficient mechanism to overcome the fundamental impediment to switch the magnetic SHE using electric fields. Such a mechanism is rooted in the polarization control of spin splitting in ferroelectric altermagnets\cite{arXiv:2411.19928,prl106802,prl106801,prl056801,nn9456,nn14618,prbL220410,prl176701} and its validity is confirmed based on the general spin-group analysis and DFT calculations on the VOI$_2$ monolayer.

\begin{figure}
\includegraphics[width=0.45\textwidth]{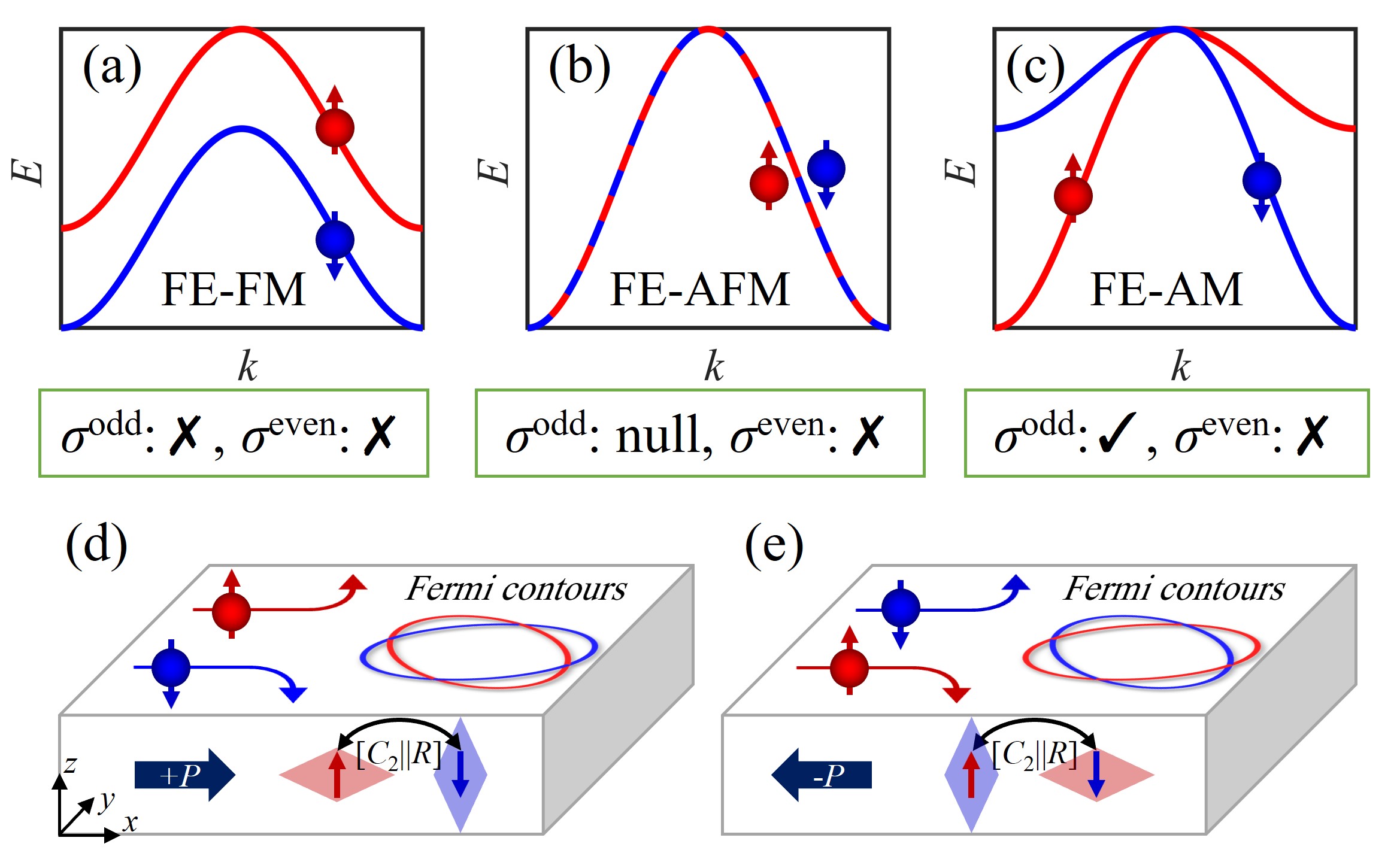}%
\caption{\label{f-1} Schematic spin splitting in a ferroelectric ferromagnet (FE-FM) (a), a ferroelectric (collinear) antiferromagnet (FE-AFM) (b), and a ferroelectric altermagnet (FE-AM) (c). The spin conductivity $\sigma^{\text{odd}}$ or $\sigma^{\text{even}}$ would be reversed (\cmark) or unchanged (\xmark) under the polarization switching. (d, e) Schematic illustration of the ferroelectric switchable magnetic SHE in ferroelectric altermagnets with positive (d) and negative (e) polarizations, as indicated by $+P$ and $-P$ respectively. The two opposite-spin sublattices are connected by the spin-group operation $[C_2||R]$, where $C_2$ and $R$ represent the spin-space and real-space operations (rotation or mirror), respectively. }
\end{figure} 

To first order in an external electric field $\bm{\mathcal{E}}$, the spin current density $\mathbf{J}_s$ (second-rank tensor) takes the form\cite{prb024410} $J_{sj}^i=\sigma_{jk}^i\mathcal{E}_k$, where a summation over repeated indices is understood. As mentioned above, $\sigma_{jk}^{i}$ contains two distinct terms i.e. $\sigma_{jk}^{i}=\sigma_{jk}^{i, \text{odd}}+\sigma_{jk}^{i, \text{even}}$. Within the linear response theory using the Kubo formalism and the relaxation time $\tau$ approximation, the $\mathcal{T}$-odd contribution $\sigma_{jk}^{i, \text{odd}}$ is\cite{prb174423,prl187204,na627,prb224401,npj46}
\begin{equation}\label{eq-1}
    \sigma_{jk}^{i, \text{odd}} = -\frac{e\hbar}{\pi}\sum_{\mathbf{k},m,n}\frac{\text{Re}[\langle\psi_{n\mathbf{k}}|\mathcal{J}^i_j|\psi_{m\mathbf{k}}\rangle\langle\psi_{m\mathbf{k}}|v_k|\psi_{n\mathbf{k}}\rangle]\Gamma^2}{[(\epsilon_F-\epsilon_{n\mathbf{k}})^2+\Gamma^2][(\epsilon_F-\epsilon_{m\mathbf{k}})^2+\Gamma^2]},
\end{equation}
and the $\mathcal{T}$-even contribution (clean limit $\Gamma\rightarrow0$) $\sigma_{jk}^{i, \text{even}}$ reads
\begin{equation}\label{eq-2}
    \sigma_{jk}^{i, \text{even}} = -2e\hbar\sum_{\mathbf{k},m\neq n}\frac{\text{Im}[\langle\psi_{n\mathbf{k}}|\mathcal{J}^i_j|\psi_{m\mathbf{k}}\rangle\langle\psi_{m\mathbf{k}}|v_k|\psi_{n\mathbf{k}}\rangle]}{(\epsilon_{n\mathbf{k}}-\epsilon_{m\mathbf{k}})^2},
\end{equation}
where $\mathcal{J}^i_j=\hbar\{\sigma_i, v_j\}/4$ is the spin current operator given in terms of the spin (velocity) operator $\sigma_i$ ($v_j$) and the reduced Planck's constant $\hbar$, $\epsilon_F$ the Fermi energy, $\Gamma$ the broadening parameter, and $\psi_{n\mathbf{k}}$ ($\epsilon_{n\mathbf{k}}$) the Bloch function (energy eigenvalue) of the $n$th band. $\tau$ and $\Gamma$ are related by $\tau=\hbar/(2\Gamma)$. In the clean limit $\Gamma\rightarrow0$, note that $\lim\limits_{\Gamma\to0}\frac{\Gamma}{\pi}\frac{1}{(\epsilon_F-\epsilon_{n\mathbf{k}})^2+\Gamma^2}=\delta(\epsilon_F-\epsilon_{n\mathbf{k}})$, Eq. (\ref{eq-1}) reduces to\cite{prb174423}
\begin{equation}\label{eq-3}
    \sigma_{jk}^{i, \text{odd}} = -e\tau\sum_{\mathbf{k}n}\langle\psi_{n\mathbf{k}}|\mathcal{J}^i_j|\psi_{n\mathbf{k}}\rangle\langle\psi_{n\mathbf{k}}|v_k|\psi_{n\mathbf{k}}\rangle\delta(\epsilon_F-\epsilon_{n\mathbf{k}}).
\end{equation}
In the case of the collinear magnetic ordering and nonrelativistic limit (e.g., altermagnets), Eq. (\ref{eq-3}) would reduce to\cite{prb224423,arxiv_tao}
\begin{equation}\label{eq-4}
    \sigma_{jk}^{i, \text{odd}} = -\frac{\hbar}{2e}(\bar{\sigma}_{jk}^{\uparrow}-\bar{\sigma}_{jk}^{\downarrow})\cos\alpha,
\end{equation}
where $\alpha$ indicates the angle between the spin polarization $i$-axis and the spin quantization axis, and the spin-resolved electrical conductivity $\bar{\sigma}_{ij}^s$ ($s=\uparrow, \downarrow$ for spin index) takes the form\cite{callaway,prb155411,prb075422}
\begin{equation}\label{eq-5}
    \bar{\sigma}_{jk}^{s}(\epsilon_F) = e^2\tau\sum_{\mathbf{k}n}v_{\mathbf{k}j}^{ns}v_{\mathbf{k}k}^{ns}\delta(\epsilon_F-\epsilon_{n\mathbf{k}}).
\end{equation}

It is evident from Eq. (\ref{eq-4}) that $\sigma_{jk}^{i, \text{odd}}$ can be switchable provided that the roles of spin-up and spin-down channels can be interchanged, which in turn swaps the spin-up and spin-down conductivities. This is typically achieved by reversing the magnetic ordering (e.g., magnetization and Néel vector) through the $\mathcal{T}$ operation. An alternative and promising approach to interchange the roles of spin-up and spin-down channels is using the dissipation-free electric field. For example, the valley-dependent spin polarization can be switched by an external electric field in two-dimensional materials due to the reversed staggered potential on the two sublattices\cite{prb161110,pra054043}. As another example, the spin texture in ferroelectrics is switchable by the electric field\cite{am509,jpd113001,rnc609}, which enables the realization of a ferroelectric spin-orbit valve effect\cite{prl076801}. Another promising demonstration of the interchange of spin-up and spin-down channels by electrical means is in ferroelectric altermagnets\cite{arXiv:2411.19928,prl106802,prl106801,prl056801,nn9456,nn14618,prbL220410,prl176701}. This is illustrated schematically in Fig. \ref{f-1}.  In a ferroelectric ferromagnet shown in Fig. \ref{f-1}(a), the polarity of spin splitting is locked to the magnetization orientation and both $\sigma^{\text{odd}}$ and $\sigma^{\text{even}}$ remains invariant upon polarization switching. In a ferroelectric (collinear) antiferromagnet shown in Fig. \ref{f-1}(b), $\sigma^{\text{odd}}$ is null due to doubly spin-degenerate bands while $\sigma^{\text{even}}$ is the same for both polarization states. As distinct from those previously considered, in a ferroelectric altermagnet shown in Fig. \ref{f-1}(c), the two real-space sublattices associated with opposite spins are connected by a rotation or mirror operation $R$\cite{prx031042,prx040501}. Along certain direction, the real-space coordination environments for spin-up and spin-down sublattices are inequivalent, which in turn results in the nonrelativistic spin splitting. In the case in which the real-space operation $A$ switching the polarization $P$ exchanges the two real-space sublattices, the spin splitting would be reversed under $A$ although the spin directions associated with the different sublattices remains unchanged. Accordingly, the spin-group operation is denoted as $[E||A]$, which can switch the polarization and the magnetic SHE simultaneously

\begin{table}
\caption{\label{table1} Non-centrosymmetric SPGs sustain the ferroelectric polarization and the altermagnetic spin splitting. Spin-group operations (not the symmetry elements of the corresponding SPG) are those to switch the polarization and the magnetic SHE simultaneously. The corresponding polarization orientation and Hamiltonian $h(\mathbf{k})$ are given in Supplemental Material, Table SI\cite{sm}.}
\begin{ruledtabular}
\begin{tabular}{cc}
SPG & Spin-group operations\\
\hline
$^2m^2m^12$  & $[E||C_{2x}], [E||C_{2y}]$ \\
$^2m^1m^22$  & $[E||C_{2x}], [E||M_z]$ \\
$^24^2m^1m$  & $[E||C_{2x}], [E||C_{2y}]$ \\
$^24^2m^1m$  & $[E||C_{2[110]}], [E||C_{2[1\bar{1}0]}]$ \\
$^14^2m^2m$  & $[E||C_{2x}], [E||C_{2y}], [E||C_{2[110]}], [E||C_{2[1\bar{1}0]}]$\\
\multirow{2}{*}{$^16^2m^2m$} & $[E||C_{2x}], [E||C_{2y}], [E||C_{2[010]}],$ \\
& $[E||C_{2[210]}], [E||C_{2[110]}], [E||C_{2[1\bar{1}0]}]$ \\
$^22$        & $[E||M_y]$ \\
$^2m$        & $[E||C_{2y}]$ \\
$^13^2m, ^26, ^26^2m^1m$ & $[E||M_z], [E||C_{2x}], [E||C_{2[110]}], [E||C_{2[010]}]$ \\
\end{tabular}
\end{ruledtabular}
\end{table}

We first analyze the electric-field switchable magnetic SHE from the general spin-group analysis. As an illustration, we consider the non-centrosymmetric SPG $^2m^2m^12$, which sustains the ferroelectric polarization and the $d$-wave altermagnetism. While the real-space operations $E, M_x, M_y, C_{2z}$ in the parent group $mm2$ render a finite polarization along the $z$ direction $P_z$, the spin-group operations $[E||E], [E||C_{2z}], [C_2||M_x], [C_2||M_y]$ enforce the $d$-wave spin splitting described by the Hamiltonian $h(\mathbf{k})=\Delta k_xk_y\sigma_z$, where $\Delta$ indicates the spin-split strength, $k_{x, y}$ are the wave numbers, and $\sigma_z$ is the Pauli spin matrix. Here, $E$ is the identity, $C_{2i}$ ($i=x, y, z$) is the twofold rotation around the $i$-axis, $M_i$ is the mirror reflection about the $i=0$ plane, and $C_2$ on the left of the double vertical bar indicates the $180^\text{o}$ rotation around an axis perpendicular to the spin. According to the transformation properties of the polarization, wave numbers and the spin, the spin-group operations $[E||C_{2x}], [E||C_{2y}]$ reverse both $P_z$ and $h(\mathbf{k})$. As a consequence, those symmetry operations swap the roles of spin-up and spin-down channels and in turn switch the $\mathcal{T}$-odd spin conductivity. This strategy is actually a consequence of the polarization switching spin splitting and so must hold in all non-centrosymmetric SPGs if the relevant symmetry operations are defined appropriately. In Table \ref{table1}, we summarize all the non-centrosymmetric SPGs that sustain both the polarization and the altermagnetism. We also list the spin-group operations that switch both the polarization and the magnetic SHE simultaneously. Those operations are not the symmetry elements of the corresponding SPG. In addition, the corresponding polarization orientation and Hamiltonian $h(\mathbf{k})$ are given in Supplemental Material, Table SI\cite{sm}. It is particulary noteworthy that the feasibility of the polarization switching under certain symmetry operation depends on the ferroelectric switching barrier, which is governed by the atomic displacements during the polarization switching. It also may be noted that the spin-group operation containing the spin-space $C_2$ operation is not taken into account although such spin-group operation may switch the polarization and $h(\mathbf{k})$ simultaneously. This is due to the fact that the spin-space $C_2$ is equivalent to the $\mathcal{T}$ operation, which is not feasible by using the electric filed.

\begin{figure}
\includegraphics[width=0.45\textwidth]{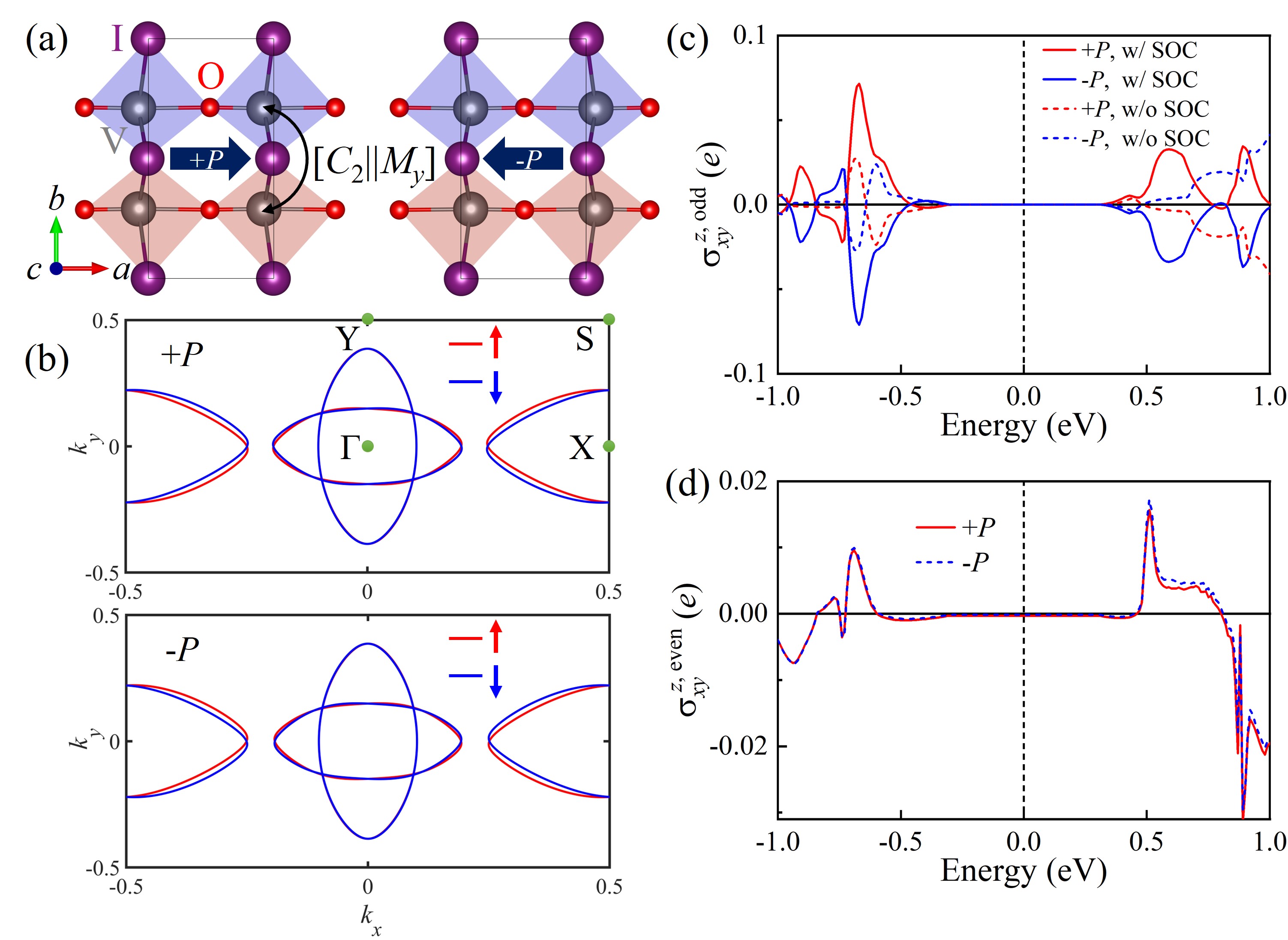}%
\caption{\label{f-2} (a) Atomic structures of the VOI$_2$ monolayer with positive ($+P$, left panel) and negative ($-P$, right panel) polarizations. The opposite-spin sublattices are connected by the spin-group operation $[C_2||M_y]$. (b) Fermi contours in the Brihouin zone of the VOI$_2$ monolayer at the Fermi energy of $0.5$ eV below the valence band maximum without SOC for positive ($+P$) and negative ($-P$) polarizations. Red and blue contour lines denote spin up and spin down channels, respectively. $k_x$ and $k_y$ are in units of $2\pi/a$ and $2\pi/b$, respectively. (c) $\mathcal{T}$-odd spin conductivity $\sigma_{xy}^{z, \text{odd}}$ as a function of energy with (w/, solid lines) and without (w/o, dashed lines) SOC for positive ($+P$, red) and negative ($-P$, blue) polarizations. (d) $\mathcal{T}$-even spin conductivity $\sigma_{xy}^{z, \text{even}}$ as a function of energy with SOC for positive ($+P$, red) and negative ($-P$, blue) polarizations.}
\end{figure} 

Having demonstrated the electric-field switchable magnetic SHE based on the general spin-group analysis, we next exemplify those phenomena in realistic altermagnets based on DFT calculations. Here we consider the dimerized VOI$_2$ monolayer, which was demonstrated as a ferroelectric altermagnet\cite{nn9456}. A VOI$_2$ monolayer crystallizes in a orthorhombic monolayer structure with the space group $Pmm2$\cite{prb195434,prbL121114,prl037203} as shown in Fig. \ref{f-2}(a). It consists of the corner-sharing (along the $a$ axis) and edge-sharing (along the $b$ axis) VO$_2$I$_4$ octahedra. A ferroelectric polarization along the $a$ axis emerges due to the shift of the V atoms from the octahedral center towards one of the bridging O atoms\cite{prb195434,prbL121114,prl037203}. We used the optimized lattice parameters $a=3.84$ and $b=7.35$ Å for DFT calculations and the other computational details are given in Supplemental Material, Sec. I\cite{sm}. From Fig. \ref{f-2}(a), the opposite-spin sublattices are connected by the spin-group operation $[C_2||M_y]$ and the corresponding SPG is $^2m^1m^22$, which sustains the $d$-wave altermagnetism. Consequently, the nonrelativistic bands are spin degenerate along the $\Gamma-X$ direction as enforced the symmetry operation $[C_2||M_y]$ or $[C_2||C_{2x}]$. This contrasts with the nonrelativistic band splitting occurring along the $\Gamma-S$ direction, as shown in Supplemental Material, Fig. S1\cite{sm}. As an illustration for the polarization switchable spin splitting, we plot the nonrelativistic Fermi contours in the Brillouin zone at the Fermi energy of $0.5$ eV below the valence band maximum in Fig. \ref{f-2}(b). It suggests a $d$-wave like spin splitting around the $\Gamma$ point. Intriguingly, the polarity of spin splitting is reversed upon polarization switching, which indicates the interconvertible roles of spin-up and spin-down channels by the polarization. This is achieved by the spin-group operation $[E||M_x]$. Moreover, the feasibility of the polarization switching can be confirmed from the kinetic pathway calculation as shown in Supplemental Material, Fig. S2(a)\cite{sm}.

\begin{figure}
\includegraphics[width=0.4\textwidth]{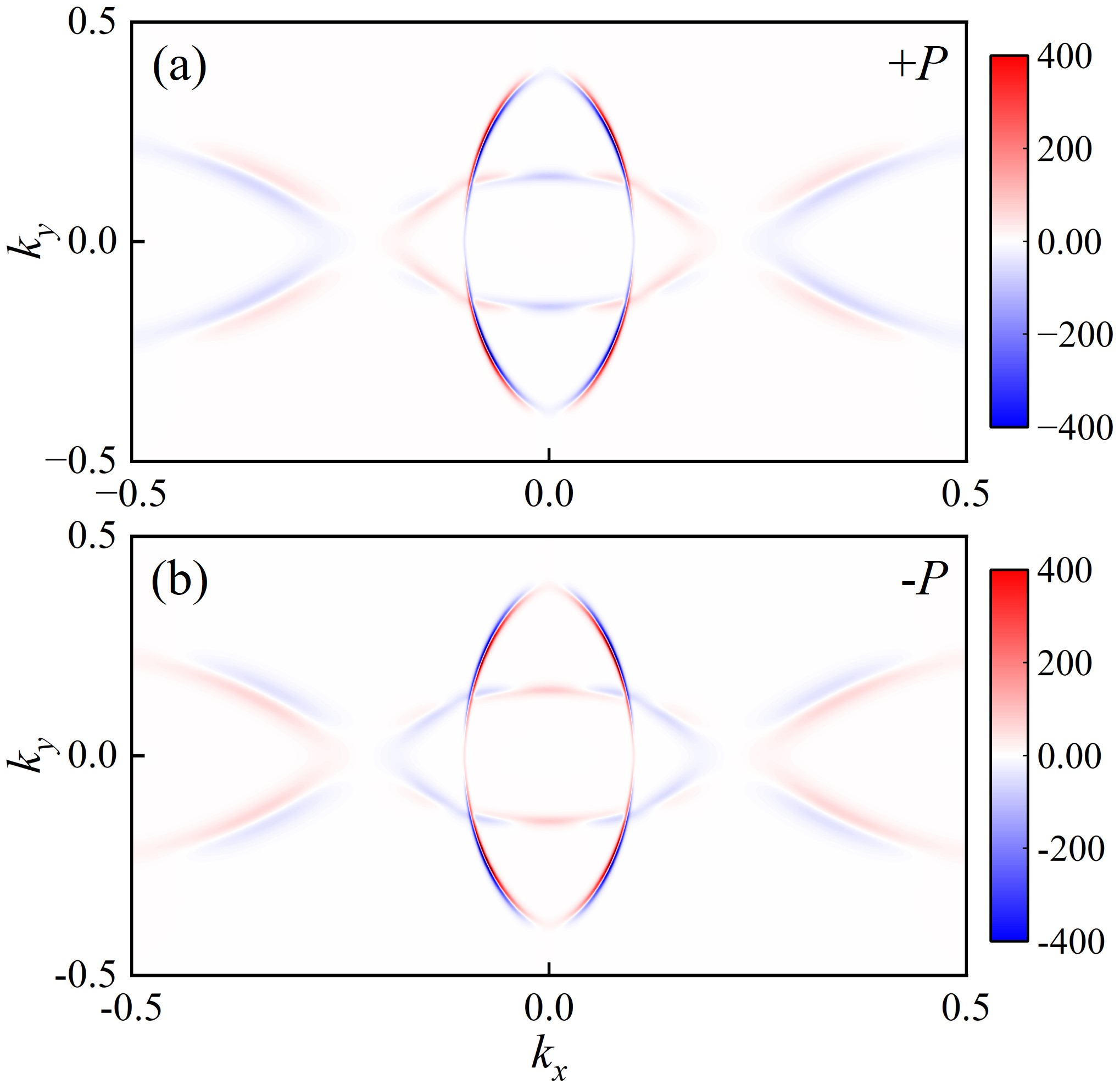}%
\caption{\label{f-3} $\mathbf{k}=(k_x, k_y)$-resolved $\sigma_{xy}^{z, \text{odd}}$ (in unit of Å$^2$) for the VOI$_2$ monolayer at the Fermi energy of $0.5$ eV below the valence band maximum without SOC for (a) positive ($+P$) and (b) negative ($-P$) polarizations. $k_x$ and $k_y$ are in units of $2\pi/a$ and $2\pi/b$, respectively.}
\end{figure} 

We now turn to the polarization control of the SHE described by the spin conductivity. Since the Néel vector is along the $c$ axis ($z$ direction) for the VOI$_2$ monolayer, we have $i=z$ and $\alpha=0$ in Eq. (\ref{eq-4}). Figure \ref{f-2}(c) shows the $\mathcal{T}$-odd spin conductivity $\sigma_{xy}^{z, \text{odd}}$ as a function of energy for positive ($+P$, red) and negative ($-P$, blue) polarizations. It clearly shows that, as expected, $\sigma_{xy}^{z, \text{odd}}$ is switchable by the polarization in both nonrelativistic and relativistic cases. This indicates the promising electric-field switchable magnetic SHE. Conversely, measuring the magnetic SHE provides a simple electrical means to detect the polarization switchable spin splitting. As distinct from $\sigma_{xy}^{z, \text{odd}}$, the $\mathcal{T}$-even spin conductivity $\sigma_{xy}^{z, \text{even}}$ remains unchanged upon polarization switching as shown in Fig. \ref{f-2}(d). This is legitimate since the $\mathcal{T}$-symmetric SHE does not depend on the polarization orientation as in non-magnetic materials. Moreover, the magnitude of $\sigma_{xy}^{z, \text{odd}}$ is about five times larger than that of $\sigma_{xy}^{z, \text{even}}$, which favors the experimental measurement of the switchable $\sigma_{xy}^{z, \text{odd}}$. Figure \ref{f-3} shows the $\mathbf{k}$-resolved $\sigma_{xy}^{z, \text{odd}}$ for the VOI$_2$ monolayer at the Fermi energy of $0.5$ eV below the valence band maximum without SOC. We see that the magnitude of $\sigma_{xy}^{z, \text{odd}}$ is more pronounced around the regions with rather small spin splitting as compared to Fig. \ref{f-2}(b). Intriguingly, the sign of the $\mathbf{k}$-resolved $\sigma_{xy}^{z, \text{odd}}$ is reversed upon polarization switching by comparison of Fig. \ref{f-3}(a) and (b).

To gain further insight, we construct a $k\cdot p$ effective Hamiltonian based on the symmetry analysis. For the VOI$_2$ monolayer with the magnetization along the out-of-plane orientation, the magnetic point group is $mm2$, which contains the symmetry elements $E$, $C_{2x}$, $M_y$ and $M_z$. Based on transformation rules of the wave vector $\mathbf{k}=(k_x, k_y)$ and Pauli matrices $\sigma_x, \sigma_y, \sigma_z$ under symmetry operations, we obtain the following $k\cdot p$ Hamiltonian around the $\Gamma$ point
\begin{equation}\label{eq-6}
    \mathcal{H}=\frac{\hbar^2k^2}{2m}+\alpha k_y\sigma_z+\Delta k_xk_y\sigma_z.
\end{equation}
The first term represents the kinetic energy given in terms of electron effective mass (isotropic approximation) $m$. The second term describes the SOC ($\alpha$ for magnitude) induced by the in-plane polarization and its validity can be confirmed by the spin textures shown in Supplemental Material, Figs. S2(b) and (c)\cite{sm}. The last term represents the $d$-wave spin splitting and $\Delta$ indicates its strength. Based on the Hamiltonian Eq. (\ref{eq-6}), we obtain the approximate expression for the $\mathcal{T}$-odd spin conductivity $\sigma_{xy}^{z, \text{odd}}$ from Eq. (\ref{eq-3}) as 
\begin{equation}\label{eq-7}
    \sigma_{xy}^{z, \text{odd}}\approx-\frac{e\tau m\Delta}{\pi\hbar^3}(\epsilon_F+\frac{3m\alpha^2}{8\hbar^2}), 
\end{equation}
where $\epsilon_F$ indicates the Fermi energy. Details of the derivation of Eqs. (\ref{eq-6}) and (\ref{eq-7}) are presented in Supplemental Material, Sec. IV\cite{sm}. We see from Eq. (\ref{eq-7}) that $\sigma_{xy}^{z, \text{odd}}$ is proportional to $\Delta$. Both the second and third terms in Eq. (\ref{eq-6}) change the signs under the spin-group operation $[E||M_x]$. Thus, the polarization reversal linked by the operation $[E||M_x]$ switches the polarity of $\sigma_{xy}^{z, \text{odd}}$ according to Eq. (\ref{eq-7}). In addition,  $\sigma_{xy}^{z, \text{odd}}$ is linearly proportional to $\epsilon_F$. This is consistent with DFT results shown in Fig. \ref{f-2}(c), which indicates that $\sigma_{xy}^{z, \text{odd}}$ increases linearly with $\epsilon_F$ around band edges.

\begin{figure}
\includegraphics[width=0.45\textwidth]{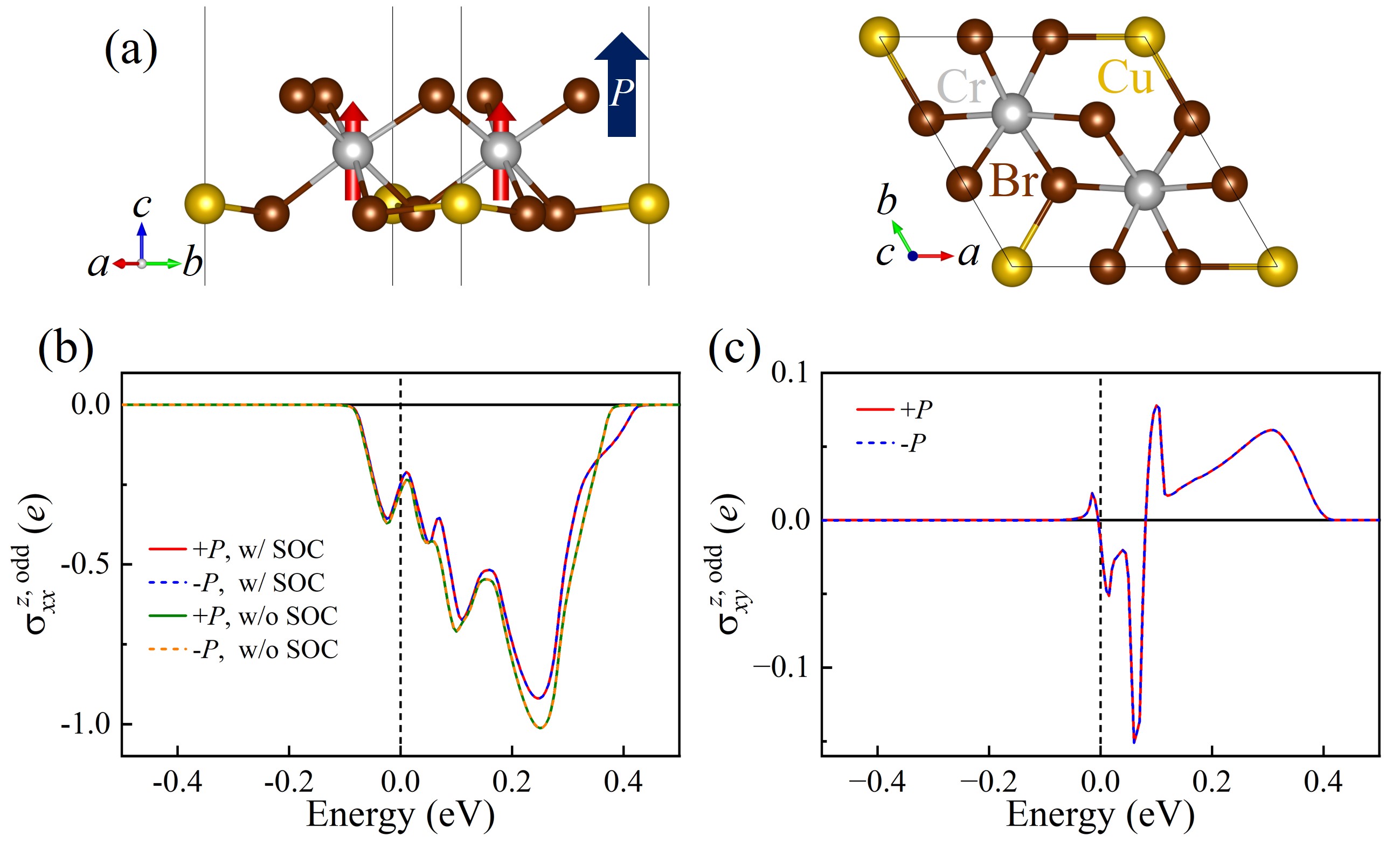}%
\caption{\label{f-4} (a) Atomic structures of the triangular Cu(CrBr$_3$)$_2$  monolayer. Red arrows denote the magnetic moments. (b) $\mathcal{T}$-odd spin conductivity $\sigma_{xx}^{z, \text{odd}}$ as a function of energy with (w/) and without (w/o) SOC for positive ($+P$) and negative ($-P$) polarizations. (c) $\mathcal{T}$-even spin conductivity $\sigma_{xy}^{z, \text{even}}$ as a function of energy with SOC for positive (a) and negative (b) polarizations, as indicated by $+P$ and $-P$ respectively. The Fermi energy has been aligned to zero.}
\end{figure} 

It is enlightening to examine the polarization control of the SHE in a ferroelectric ferromagnet. Without loss of generality, we consider the Cu(CrBr$_3$)$_2$ monolayer, which was demonstrated as a ferroelectric ferromagnet\cite{npj180}. A Cu(CrBr$_3$)$_2$ monolayer crystallizes in a triangular structure with the space group $P31m$ as shown in Fig. \ref{f-4}(a). We used the lattice constants of $a=b=6.67$ Å for DFT calculations and the other computational details are given in Supplemental Material, Sec. I\cite{sm}. The polarization along the $c$ axis (out-of-plane) is induced by the Cu decoration at the center of the Cr honeycomb lattice, which results in unbalanced positive and negative charge centers\cite{npj180}.

Given the fact that the magnetic easy axis for the Cu(CrBr$_3$)$_2$ monolayer is along the $c$ axis ($z$ direction), we have $i=z$ and $\alpha=0$ in Eq. (\ref{eq-4}). Figure \ref{f-4}(b) shows the $\mathcal{T}$-odd spin conductivity $\sigma_{xx}^{z, \text{odd}}$ as a function of energy for positive ($+P$) and negative ($-P$) polarizations. Since the Cu(CrBr$_3$)$_2$ monolayer reveals half-metallicity around the Fermi energy (see Supplemental Material, Fig. S2\cite{sm}), $\sigma_{xx}^{z, \text{odd}}$ is negative inferred from Eq. (\ref{eq-4}). As distinct from the VOI$_2$ monolayer shown in Fig. \ref{f-2}(c), $\sigma_{xx}^{z, \text{odd}}$ is independent of the polarization orientation in both nonrelativistic and relativistic cases. This is due to fact that the polarity of spin splitting caused by the exchange coupling is locked to the magnetization orientation. Consequently, the time-reversal operation is required to switch the $\mathcal{T}$-odd spin conductivity in ferromagnets. A similar result is also observed for the $\mathcal{T}$-even spin conductivity $\sigma_{xy}^{z, \text{even}}$ as shown in Fig. \ref{f-4}(c). It is seen that both the $\mathcal{T}$-odd and $\mathcal{T}$-even spin conductivities remain unchanged upon polarization switching consistent with Fig. \ref{f-1}(a).

In summary, we show that the deterministic and nonvolatile switching of the magnetic SHE using electric fields can be achieved in ferroelectric altermagnets. It works in such a way that the electric-field-induced polarization reversal switches the polarity of the spin splitting proportional to the $\mathcal{T}$-odd spin conductivity. Although we exemplify this in two-dimensional ferroelectric altermagnets due to the computational demand, the proposed mechanism based on the general spin-group analysis can be extended to be applicable to bulk ferroelectric altermagnets. Our work paves the practical way to design electrically programmable spintronic devices based on the switchable magnetic SHE.

This work was supported by the National Natural Science Foundation of China (No. 12274102 and No. 52372004). The atomic structures were produced using the VESTA software\cite{vesta}.

\emph{Data availability}—The data that support the findings of this article are not publicly available. The data are available from the authors upon reasonable request.

\end{document}